# Magnetoresistance and Spin-Transfer Torque in Magnetic Tunnel Junctions

J. Z. Sun[1] and D. C. Ralph[2]

[1]IBM T. J. Watson Research Center, Yorktown Heights, New York 10598, USA

[2]Laboratory of Atomic and Solid State Physics, Cornell University, Ithaca, New York 14853, USA

Abstract:

We comment on both recent progress and lingering puzzles related to research on magnetic tunnel junctions (MTJs). MTJs are already being used in applications such as magnetic-field sensors in the read heads of disk drives, and they may also be the first device geometry in which spin-torque effects are applied to manipulate magnetic dynamics, in order to make nonvolatile magnetic random access memory. However, there remain many unanswered questions about such basic properties as the magnetoresistance of MTJs, how their properties change as a function of tunnel-barrier thickness and applied bias, and what are the magnitude and direction of the spin-transfer-torque vector induced by a tunnel current.

Magnetic tunnel junctions (MTJs) present many interesting scientific questions that are important for applications. Recent breakthroughs in the field include the demonstration of spin-transfer-driven magnetic switching in MTJs [1,2] and very large, room-temperature tunnel magnetoresistance [3,4,5,6]. These advances have attracted attention because they show that spin-transfer torque is a promising mechanism for achieving the controlled manipulation of magnetic moments in tunnel-junction devices such as nonvolatile memory elements, and that the tunnel magnetoresistance is large enough for many memory and field-sensing applications. From a basic-science point of view, spin-transfer experiments with MTJs are also interesting because they may be able to provide an independent means of measuring the flow of spin angular momentum, complementary to magneto-resistance measurements, thus giving new insights into long-standing mysteries about the mechanisms that affect spin-polarized tunneling. In this article, we will attempt to summarize briefly some of the current state of understanding in this field, and to highlight interesting open questions that may be productive areas for future research.

Tunnel Magnetoresistance and its Bias Dependence

The basic phenomenon of spin-dependent tunneling in solid-state device structures has been known for more than 30 years, starting from the work of Julliere [7], whose experiments also established a quantitative method for measuring the spin-polarization of a tunnel junction containing ferromagnetic electrodes. Recent research into spin dependent tunneling in transition-metal-based MTJs has resulted in tunnel magneto-resistances (TMRs) that have surpassed 500% at room temperature [8].

Conventionally, TMR is defined as the ratio $[R_{AP}-R_P]/R_P$, where $R_{AP}$ is the differential resistance for an antiparallel orientation of the magnetic moments in the two electrodes and $R_P$ is the differential resistance for parallel moments. Junctions based on spin-dependent tunneling are now used commercially in computer hard-drive read-heads and magnetic random access memory chips. For these types of applied devices, several types of tunnel barriers have been studied extensively: $AlO_x$ [9, 10], $TiO_x$, $HfO_x$ and ZrO [11,12], and more recently, MgO [3-6]. Exploratory activities have also included a much wider range of tunnel barrier materials, such as AlN [13], $SrTiO_3$ [14,15,16], and $LaAlO_3$ [17].

For well-controlled materials systems, such as FeCo electrodes interfaced with high-quality $AlO_x$ tunnel barriers, the magnitude of MR at low junction bias for many tunnel junction systems can be understood reasonably well within the framework of the Meservey-Tedrow model [18]. The spin-polarization factor $P_s$ derived from ferromagnet / insulator / superconductor (F/I/S) tunnel junctions [19] is in general agreement with that derived from the TMR of ferromagnet / insulator / ferromagnet (F/I/F) junctions for tunnel barriers prepared under similar conditions. However, this type of quantitative comparison may not always be straightforward, especially for crystalline barrier and electrode interfaces such as Fe(100)/MgO(100). In such systems the F/I/F and F/I/S junctions may differ in important materials properties such as orientation and texture [20]. Consequently, the tunneling probability may depend on the orientation of the electron wave vector in ways that differ between the F/I/S and F/I/F junctions, making the analysis and comparison of effective spin polarization factors more complex and challenging.

One essentially-universal feature of MTJs is a significant reduction in the TMR as a function of increasing bias. For MTJs made from simple transition-metal magnets, usually the TMR drops to half of the zero-bias value for a *dc* bias between 0.1 and 0.5 V. One such example is given in Figure 1. As illustrated by this device, the bias dependence is ordinarily stronger for the antiparallel-moment (AP, or high resistance) state than for the parallel (P, low resistance) state. Strongly bias-dependent decreases of TMR are also observed in all-oxide MTJs involving both nearly half-metallic systems (La(Sr/Ca)MnO$_3$-SrTiO$_3$- La(Sr/Ca)MnO$_3$ [21] and Co-SrTiO$_3$-La$_{0.7}$Sr$_{0.3}$MnO$_3$ [16]) and very-high-quality Heusler alloy tunnel junctions [22,23]. One proposed mechanism for the decrease in TMR as a function of bias is interface magnon excitation [24]. Another that is sometimes invoked is energy dependence in the electronic band structure of the magnetic electrodes [25,26]. A third potential mechanism is the presence of defect states within the tunnel barrier, which may allow an increasing amount of defect-state-assisted tunnel current with increasing junction bias. This type of inelastic tunnel current could effectively dilute the spin-polarization of the tunnel current, thereby explaining the decrease in TMR at elevated bias [27,28,29]. However, for the type of MgO-based MTJ shown in Fig. 1, the fact that the conductance difference $G_P - G_{AP}$ decreases significantly at elevated bias suggests that the mechanism must act somewhat differently for parallel and antiparallel magnetic orientations -- a simple opening of additional spin-independent conduction channels at elevated bias would not by itself result in a change in $G_P - G_{AP}$.

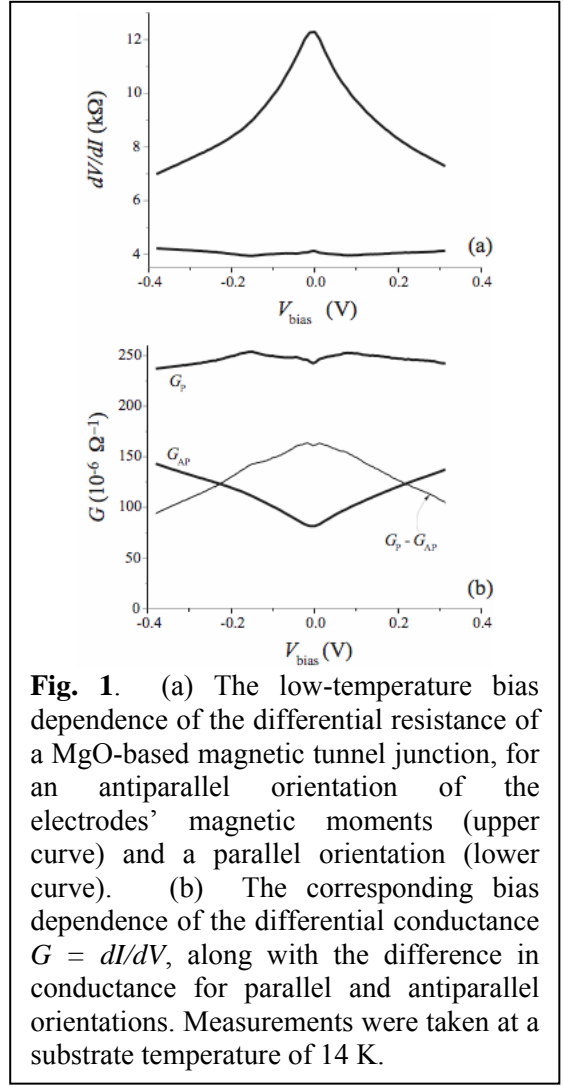

**Fig. 1**. (a) The low-temperature bias dependence of the differential resistance of a MgO-based magnetic tunnel junction, for an antiparallel orientation of the electrodes' magnetic moments (upper curve) and a parallel orientation (lower curve). (b) The corresponding bias dependence of the differential conductance $G = dI/dV$, along with the difference in conductance for parallel and antiparallel orientations. Measurements were taken at a substrate temperature of 14 K.

The uncertainty regarding the mechanisms affecting the bias dependence of the TMR has motivated careful studies of the electronic structure of thin oxide tunnel barriers and what defects they contain. For both amorphous $AlO_x$ and crystalline MgO barriers, it was found experimentally that a significant concentration of electronic trap states can exist in an as-prepared tunnel barrier. Upon vacuum annealing above 350 °C a majority of the trap states can be eliminated, and the tunnel barriers show much more ideal tunnel characteristics as a function of bias [30]. Another set of studies [31] focused on the effects of interface-mediated resonance states [32] and also states associated with carbon contamination at interfaces in Fe/MgO/Fe epitaxial junctions. They found that carbon contamination can cause a dramatic increase in the bias dependence of the TMR, or even a change of sign of the TMR at elevated bias.

Despite considerable effort, no unambiguous correlation has been established between the bias-dependence of TMR and any of the mechanisms proposed to explain the effect. The experimental data to date have been indirect and suggestive, at best, of the specific mechanisms at play. The subject is challenging in part because of the difficulties in making direct measurements of physical phenomena buried within a tunnel barrier, such as interface magnon excitation and/or resonant tunneling via defect states. Clearly, additional study is warranted into the details of the magnetic, electronic, and chemical structure of MTJs and how all of these affect TMR.

Dependence of Resistance and TMR on Barrier Thickness

The dependence of a MTJ's TMR and resistance-area product (RA) on tunnel-barrier thickness is a subject of both theoretical and practical interest. Several recent results present interesting puzzles. A basic ballistic tunneling model would predict that the tunnel current (at constant voltage) should decay exponentially as a function of increasing barrier thickness. This is indeed observed experimentally in many cases. However, an exponential relationship between a junction's RA and barrier thickness is not necessarily indicative of simple ballistic tunneling. This is because in any practical solid-state tunnel-junction structure the barrier is unlikely to be atomically smooth and defect-free over the junction area. A distribution of tunnel barrier thicknesses and the presence of defects may significantly alter the tunnel process, and at the same time give an apparent exponential decay of tunnel conductance which is not directly related to the true barrier height or its corresponding direct tunnel probability [33,14]. In addition, the effective electron mass of the tunnel state within the barrier may become a variable, too, due to a two-band effect if the Fermi-level approaches the barrier's band-gap center [34], as is expected to be the case for Fe/MgO/Fe MTJs based on band-structure calculation [3]. One therefore needs to be cautious with parameters obtained by fitting current-voltage curves to simple free-electron, rectangular-barrier tunneling models that are sometimes used to estimate tunnel-barrier heights (Brinkman-Dynes or Simmons models [35]).

The dependence of TMR on barrier thickness, and hence on junction RA, has been particularly interesting for epitaxial MTJs such as those in Fe(100)/MgO/Fe(100) devices. In such samples, a model calculation indicates that the anti-parallel tunnel matrix element is close to zero for tunneling electrons traveling purely perpendicularly to the interface, due to band-structure symmetries [3,4]. As a tunnel barrier becomes thicker, the forward-momentum selection should become more stringent and the AP conduction paths should be blocked more effectively, thereby producing a larger and larger value of TMR, eventually approaching saturation only at TMR values well in excess of 15,000% [36]. Experimentally, Yuasa *et al.* [6] did find that the TMR of such junctions generally increases with increasing MgO barrier thickness. However the amount of increase is more modest than predicted, and it saturates for barrier thicknesses beyond about 2 nm. This has been attributed in part to the presence of interface defect states and their role in coupling perpendicular-momentum states to states with non-zero parallel momentum [37]. This coupling would generate a residual conductance and limit the amount of reduction in tunnel conductance for AP state, and hence limit the increase of MR upon increasing barrier thickness. Other scattering sources such as trap states or misfit dislocations within the barrier would presumably behave in a similar way to keep the tunnel resistance in the AP state from diverging. It should be noted that even epitaxial Fe/MgO/Fe junctions are far from defect-free -- they have a significant density of misfit dislocations in the barrier because there is a very substantial lattice mismatch between Fe and MgO [6]. It remains a very interesting open question what are the mechanisms by which barrier defects may affect the TMR, and also whether different types of defect structures might produce qualitatively different consequences. In this regard, it is interesting to note that MgO barriers deposited by sputtering can have very large TMR values similar to barriers made by molecular beam epitaxy, despite the fact that the sputtered films are significantly more disordered, yet the ultimate TMR values are quite sensitive to annealing protocols which can manipulate the defect structures [5,6]. A detailed understanding the defect structures in MgO barriers is still being developed. For sputter-deposited junctions, experimental evidence points towards the importance of grain-to-grain epitaxy, and a non-uniform defect distribution around columnar grains that are coherent across the MgO barrier [38].

In addition to the overall increase with increasing barrier thickness, the TMR of epitaxial Fe/MgO/Fe (100) junctions has been found to exhibit oscillations as a function of MgO barrier thickness. This was first reported by Yuasa *et al*. [6,39]. The phenomenon is still not well understood. A recent report [39] associates the oscillations primarily with the parallel state of the junction resistance, which is counter-intuitive, because one would expect well-defined oscillations to occur only when tunneling is concentrated near localized regions of momentum space (so-called "hot-spots") and this sort of band-structure restriction on tunneling should be more significant for the anti-parallel state [40]. More recently, oscillations of TMR ratio as function of MgO thickness have also been reported for Heusler alloy-based MTJs of the structure CCFA/MgO/Co$_{50}$Fe$_{50}$ [41,42,43,44,45 and references therein], where CCFA = Co$_2$Cr$_{0.6}$Fe$_{0.4}$Al is the Heusler alloy. These show the same period reported earlier by Yuasa *et al*. [39], although in the case of Heusler alloy studies the oscillations were present for both the parallel and the anti-parallel states. The oscillation period in both cases was about 3.1 Å, significantly different than the lattice constant of MgO. This observation was used as an argument against the cause of the oscillations being any growth-induced periodicity in roughness.

The fabrication of very thin (low RA) tunnel junctions with good TMR is an important challenge for applications, in order to minimize the Johnson noise in read heads and the bias voltage necessary to excite spin-torque-driven dynamics in memory devices and oscillators. In devices made to date, the TMR decreases significantly in very thin barriers for both MgO junctions (typically < 1 nm) [46,47] and AlO$_x$ junctions [48]. The mechanisms involved can be complex, but this decrease depends sensitively on thin film processing conditions. Therefore, the leading cause for the rapid decrease of TMR for very thin barriers is probably related to barrier defects such as pin-holes [49].

The Spin-Transfer-Torque Vector in Tunnel Junctions

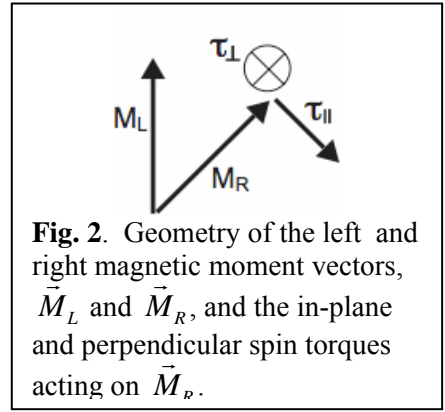

**Fig. 2**. Geometry of the left and right magnetic moment vectors, $\vec{M}_L$ and $\vec{M}_R$, and the in-plane and perpendicular spin torques acting on $\vec{M}_R$.

In order to present a simple introduction to the physics of spin-transfer torque in a MTJ, we will make the approximation that for a junction with a sufficiently small area, the magnetic moments in the two electrodes can each be treated as simple single-domain macrospins, $\vec{M}_R$ and $\vec{M}_L$. This is, of course, an assumption that will break down in many experimental configurations, in which case a more complex micromagnetic picture is required. Because the spin-transfer torque is expected to be zero if $\vec{M}_R$ and $\vec{M}_L$ are parallel, we will consider a general, non-collinear geometry with an angle $\theta$ between the moments (Fig. 2). We will also assume that the exchange interactions within each material are sufficiently strong that each of the magnetization vectors can be considered to have a fixed magnitude, meaning that the spin torque on a magnet caused by an interaction with conduction electrons will act only perpendicularly to its magnetization vector. The spin-torque vector $\vec{\tau}$ acting on each magnet can therefore be decomposed into two components, one perpendicular to the plane defined by $\vec{M}_R$ and $\vec{M}_L$, and one within this plane. For definiteness, we will consider the torque acting on the right-hand moment $\vec{M}_R$.

Even in the absence of any bias applied across the tunnel junction, there can be a conduction-electron-mediated torque on $\vec{M}_R$ acting in the direction perpendicular to the plane defined by $\vec{M}_R$ and $\vec{M}_L$ [50,51]. This is often called the "interlayer exchange coupling" or "conservative exchange coupling" because it is in the same direction as the energy-conserving torque on a magnetic moment due to exchange coupling with another moment. For simplicity, we will just refer to it as the perpendicular torque component or $\tau_\perp$. It can be considered a consequence of spin angular momentum transferred to a magnetic layer by all the conduction electrons with energies below the Fermi energy, which are incident on the tunnel junction from both the left and right sides [50]. It is analogous to the so-called RKKY interaction for magnetic multilayers with normal-metal spacers [52]. This equilibrium torque in a tunnel junction is of considerable interest by itself due to recent measurements as a function of barrier thickness that found unexpected results, and because of theoretical predictions that it will be affected strongly by impurity states in the barrier [53,54,55]. For symmetric tunnel junctions (with the same electrode materials on the left and right), when a bias voltage $V$ is applied, theory predicts that the perpendicular torque will not have any dependence to first order in $V$, so that for sufficiently low bias the dependence should go as $\tau_\perp(V) = A_0 + A_1 V^2$, where $A_0$ and $A_1$ are constants [51,56]. The lack of a term linear in $V$ can be understood from the result that, for a symmetric tunnel junction, the contribution to the perpendicular torque on $\vec{M}_R$ from an electron incident from the left of the junction at the Fermi level has both the same magnitude and the same sign as the torque due to an electron incident from the right at the Fermi level [sections III-IV of 50]. Therefore, when a small bias is applied, causing more electrons to be incident on the tunnel junction from one side and fewer from the other, to first order the total perpendicular torque should not change. The dependence proportional to $V^2$ can arise from the presence of any energy dependence in the transmission factor across the junction. This will cause the perpendicular torque contributed by electrons incident from the left at an energy $\varepsilon_F + eV/2$ to differ from the contribution of electrons incident from the right at an energy $\varepsilon_F - eV/2$, thereby changing the total perpendicular torque after the sum over all occupied states is completed.

The other component of the spin-torque on $\vec{M}_R$, the one in the plane defined by $\vec{M}_R$ and $\vec{M}_L$, is expected to be by far the dominant *bias-dependent* component, at least for small bias, because its magnitude is predicted to be proportional to $V$ for small bias. We will call this the in-plane torque, or $\tau_\parallel(V)$. It was first predicted by Slonczewski [50]. In discussing the bias dependence of the torque, it is often more convenient to consider the derivative of the torque, $d\vec{\tau}/dV$, rather than the torque itself, and we will do so frequently below. This derivative quantity has been given the name differential "torkance" [57], in analogy to the differential conductance $dI/dV$. When the spin-torque is roughly proportional to $V$, the torkance is roughly constant and its magnitude conveys the effectiveness with which a bias generates the torque. When the torque is not exactly proportional to $V$, the bias dependence of the torkance highlights the deviations. The torkance is particularly useful for analyzing the most-accurate experiments for measuring the spin torque, spin-transfer-driven ferromagnetic resonance (ST-FMR, discussed below), because this technique provides a determination of $d\vec{\tau}/dV$ rather than the torque itself.

The in-plane component of the torkance ($d\tau_{\parallel}/dV$) can be calculated using a simple geometric construction introduced by Slonczewski, that is equally applicable to either all-metal multilayers or tunnel junctions [51,57]. If one calculates the current at a location well within a ferromagnet, it cannot have a spin component in any direction other than the direction of the magnetic-moment vector [58], so that the total spin current flowing within, e.g., the right-hand magnetic electrode is $(I_{R,+} - I_{R,-})\hbar\hat{r}/(2e)$, where $I_{R,+}$ and $I_{R,-}$ are the majority (spin up) and minority (spin down) spin currents and $\hat{r}$ is a unit vector in the direction of $\vec{M}_R$. The expression for spin flow within the left-hand electrode is analogous. Assuming that during tunneling no angular momentum is lost to any processes other than transfer to $\vec{M}_R$ and $\vec{M}_L$, the sum of the bias-dependent parts of the spin torque vectors acting on both magnetic electrodes can be calculated as just the net angular momentum transferred from the current as it passes through the tunnel junction:

$$\vec{\tau}_L + \vec{\tau}_R = \frac{\hbar}{2e}\left[(I_{R,+} - I_{R,-})\hat{r} - (I_{L,+} - I_{L,-})\hat{l}\right]. \tag{1}$$

This holds for both the in-plane and perpendicular components of the torque. The in-plane component of torque on $\vec{M}_R$ can be calculated easily by taking the dot product of this equation with $\hat{l}$ and dividing by $\sin(\theta)$ (since $\hat{l} \cdot \vec{\tau}_L = 0$ and $\hat{l} \cdot \vec{\tau}_R = -\sin(\theta)\tau_{R,\parallel}$):

$$\tau_{R,\parallel} \equiv \tau_{\parallel} = \frac{\hbar\left[(I_{L,+} - I_{L,-}) - (I_{R,+} - I_{R,-})\cos(\theta)\right]}{2e\sin(\theta)}. \tag{2}$$

For the case of a tunnel junction (assuming that the resistance of the leads is negligible compared to the tunnel-junction itself), the majority and minority spin currents can be evaluated by Fermi's golden rule and, assuming that the tunneling mechanism does not depend on spin operators explicitly [57], they can be expressed in terms of conductance amplitudes $G_{\sigma\sigma'}$ between spin channels ($\sigma, \sigma' = \pm$ are spin indices for the left and right electrodes) [51]:

$$\begin{aligned}
dI_{L,+}/dV &= G_{++}\cos^2(\theta/2) + G_{+-}\sin^2(\theta/2) \\
dI_{L,-}/dV &= G_{-+}\sin^2(\theta/2) + G_{--}\cos^2(\theta/2) \\
dI_{R,+}/dV &= G_{++}\cos^2(\theta/2) + G_{-+}\sin^2(\theta/2) \\
dI_{R,-}/dV &= G_{+-}\sin^2(\theta/2) + G_{--}\cos^2(\theta/2).
\end{aligned} \quad (3)$$

Taking the derivative of Eq. (2) with respect to bias and substituting Equations (3) then gives a simple expression for the magnitude of the in-plane component of the torkance $d\tau_\parallel/dV$ acting on $\vec{M}_R$ [51,57]:

$$\frac{d\tau_\parallel}{dV} = \frac{\hbar}{4e}(G_{++} - G_{--} + G_{+-} - G_{-+})\sin(\theta). \quad (4)$$

The sign of the torque on $\vec{M}_R$ is, for predominantly majority-electron tunneling, to force $\vec{M}_R$ toward the direction of $\vec{M}_L$ for electron flow from the left to right. The differential conductances associated with charge flow for the parallel and antiparallel magnetic configurations of a MTJ can be written analogously

$$\left(\frac{dI}{dV}\right)_P = G_{++} + G_{--}, \quad \left(\frac{dI}{dV}\right)_{AP} = G_{+-} + G_{-+}. \quad (5)$$

These expressions (Equations (3)-(5)) appropriate for tunnel junctions are different from those for an all-metal multilayer because in that case the calculation of spin currents must take into account multiple reflections of electron waves from the interfaces of the multilayer. Also, all-metal junctions are usually of much lower resistance, making effects related to spin accumulation more important, and this can cause the spin-transfer torque to depend not only on the local interface properties but on the full structure of the sample and the electrodes within a spin-diffusion length of $\vec{M}_R$. These factors lead to a more complicated dependence on the angle $\theta$ between the magnetic moments than the simple sinusoidal terms present in Equations (3)-(5) [59,60,61]. However, the simple expressions for the tunnel junction case enable the prediction of a particularly direct relationship between the spin-transfer torkance and the electrical conductance of an MTJ. For a symmetric tunnel junction and for small biases one should expect $G_{+-} \approx G_{-+}$, and for a tunnel junction with large TMR, $G_{++} \gg G_{--}$. Consequently, the voltage derivative of the in-plane torque at any angle $\theta$ is predicted to be simply proportional to the differential conductance for parallel electrodes, times the factor $\hbar \sin(\theta)/(4e)$. More quantitatively, the term $G_{--}$ can be taken into account with the prediction that at $V=0$

$$\frac{d\tau_{\parallel}}{dV}(V=0) = \frac{\hbar}{4e} \frac{2P_S}{1+P_S^2} \sin(\theta) \left(\frac{dI}{dV}\right)_P, \qquad (6)$$

where $P_S$ is the tunneling polarization [57].

As a function of the angle $\theta$ between the left and right magnetic moment directions, the tunneling model predicts that $d\tau_{\parallel}/dV$ has a simple $\sin(\theta)$ dependence. Since $|\sin(\pi-\theta)| = |\sin(\theta)|$, this implies that the strength of the torque per unit applied *voltage* for a given MTJ should be similar for small deviations from parallel and antiparallel magnetic alignment. However, because high-quality tunnel junctions have very different resistances in the parallel and antiparallel states, the model predicts a distinct asymmetry in the strength of the torque per unit *current*. The torque per unit current should be significantly stronger near the high-resistance antiparallel state than near the low-resistance parallel state. This is indeed reflected experimentally in that the critical current for magnetic reversal in high-TMR tunnel junctions is generally lower for antiparallel-to-parallel switching than for parallel-to-antiparallel [62], when the magnetic free layer experiences zero total magnetic field. The critical voltages for switching in high-TMR MTJs are generally much closer to each other, although they can differ slightly due to effects of heating and magnetic-field nonuniformities that can vary between the two magnetic states.

In the first generation of experiments to observe spin-torque-driven excitations in all-metal multilayers and MTJs, it was difficult to make a quantitative analysis of the torque because the only type of excitation that could be studied in detail was switching. The critical bias for spin-torque-driven switching is affected by thermal fluctuations (that can reduce it significantly below its intrinsic zero-temperature value), and non-single-domain-like excitations can also cause significant deviations from simple "macrospin" models. Often the theoretical prediction was for an instability threshold in the small-amplitude, linearized magneto-dynamics, whereas experimental observations of switching occur at a different, significantly-larger bias threshold. In such cases numerical simulations have been used to indicate semi-quantitative agreements between model assumptions and experimental observations. However, even in the best cases, the thresholds for spin-torque driven excitations are determined by a competition between the spin torque and magnetic damping. Because no reliable technique had existed to determine the damping in individual magnetic nanostructures, this prevented any very precise determination of the spin-transfer torque from measurements of switching thresholds.

Truly quantitative measurements of the magnitude and direction of the spin-transfer torque vector, and also the magnetic damping, have recently become possible in individual nanoscale MTJs, with the development of the technique of spin-transfer-driven ferromagnetic resonance (ST-FMR) [63,64,65,66]. In this type of experiment, the initial orientations of $\vec{M}_R$ and $\vec{M}_L$ are offset by an exchange bias and/or an applied magnetic field, and spin transfer from a microwave-frequency current is used to apply an oscillating torque. When the applied frequency matches the resonance frequency of one of the magnetic normal modes, the resulting precession causes the resistance to oscillate at the same frequency as the applied current, and mixing between these resistance oscillations and the microwave current produces an easily-measurable *dc* voltage signal. The linewidth of this resonance provides a measure of the magnetic damping. As first pointed out by Tulapurkar *et al*. [63], the resonance magnitude and lineshape as a function of frequency allow a determination of the magnitude and direction of the bias-dependent part of the spin torque. If only an in-plane component $d\tau_{\|}/dV$ is present, the lineshape is predicted to be a simple Lorentzian that is symmetric in frequency about the center of the resonance ($S(\omega) = 1/[1+((\omega-\omega_0)/\Delta)^2]$, where $\omega_0$ is the center frequency and $\Delta$ is the linewidth). The *dc* voltage on resonance has one sign for precession of the left magnetic moment and the opposite sign for precession of the other moment, allowing the nature of the magnetic normal modes to be distinguished. If in addition (or instead) there is a non-zero perpendicular component $d\tau_{\perp}/dV$, this adds a voltage signal that has a shape of a frequency-antisymmetric Lorentzian ($A(\omega) = ((\omega-\omega_0)/\Delta)S(\omega)$). By fitting the resonance shape, it is straightforward in MTJs to measure the amplitudes of both the frequency-symmetric and frequency antisymmetric parts, and from these to determine the magnitudes of both the in-plane and perpendicular components $d\tau_{\|}/dV$ and $d\tau_{\perp}/dV$. (It is important to note that if the precession axis is not along a high symmetry axis of the magnetic anisotropy, a frequency-antisymmetric component can appear even in the absence of a perpendicular torque [67,68], but this prediction should not affect the experiments that we will discuss.) ST-FMR can also be performed in all-metal multilayers as well as MTJs, but the results to date for the strength of the spin torque in metallic devices are not yet quantitative because there can be an additional spin-emission contribution to the ST-FMR signal in all-metal devices that is negligible for tunnel junctions [68].

At this time, ST-FMR measurements on CoFeB/MgO/CoFeB tunnel junctions have been performed by two groups, and their interpretations do not fully agree. A collaboration from Tsukuba, the Canon ANELVA Corp, and Osaka (Tulapurkar *et al.* [63]) found resonances with lineshapes having a significant frequency-antisymmetric component even for zero applied bias across the tunnel junction, from which they initially concluded that there is a significant perpendicular component $d\tau_{\perp}/dV$ even near *V*=0. With our collaborators [65], we measured ST-FMR resonances that are frequency-symmetric Lorentzians at *V*=0, with no frequency-antisymmetric component to within the accuracy of our apparatus. This led us to conclude that $d\tau_{\perp}/dV$=0 at *V*=0, so that $\tau_{\perp}(V)$ has no bias-dependent part proportional to *V* at low bias in our symmetric tunnel junctions. This is in agreement with the theory described above [51,56]. At this time, the reason for the discrepancy between experiments is not clear, and additional measurements are certainly warranted, to explore different materials, tunnel-barrier thicknesses, and device geometries. Recently, the Tsukuba-ANELVA-Osaka collaboration has reported new measurements in which the ST-FMR peaks are symmetric at *V*=0, and have suggested that the asymmetries observed previously can arise when samples have magnetic states that are not spatially uniform [66]. This seems reasonable, since the Tsukuba-ANELVA-Osaka collaboration has used samples with larger cross sections than ours (100 nm × 200 nm and 70 nm × 250 nm, as compared to our 50 nm × 100 nm and 50 nm × 150 nm samples), which may be more prone to the formation of non-uniform states. Another difference between the experiments was in the sample design -- our group used devices in which the milling step that defined the device cross section stopped at the tunnel barrier, so that the magnetic "fixed layer" in the bottom electrode remained connected to an extended film. The bottom electrode in the devices of Tulapurkar *et al.* was etched through, giving it a cross section similar to the top electrode. In our geometry, there should be no significant excitation of magnetic precessional modes within the bottom electrode because of both strong exchange coupling to the extended film and weak magnetic dipole coupling between the top and bottom magnetic layers. However, the samples examined by Tulapurkar *et al.* may have supported fixed-layer modes or coupled modes involving precession in both layers, which might complicate interpretation of the resonant peak shapes.

In addition to experiments at zero voltage, the spin-transfer torque vector in MgO tunnel junctions has been measured as a function of bias [65,66]. Both our group and the Tsukuba-ANELVA-Osaka collaboration found that the vector $d\vec{\tau}/dV$ rotates out of the plane as *V* is changed from 0, so that at large bias there can be a significant perpendicular component.  These measurements indicate, in agreement with the predictions, that $d\tau_\perp/dV \propto V$ at low bias, so that $\tau_\perp(V)$ has the form $A_0 + A_1 V^2$.  For the in-plane component, our group found that $d\tau_\parallel/dV$ at V=0 is equal to the magnitude predicted by Eq. (6) within our measurement accuracy (±15%).  Also, $d\tau_\parallel/dV$ is approximately constant as a function of bias over a large range, varying by less than 8% for -300 mV < V < 300 mV (consistent with a previous measurement [69]).  This lack of bias dependence in excellent agreement with our expectations (based on Eq. (4) and Eq. (5)) that $d\tau_\parallel/dV \propto (dI/dV)_P$ because the parallel conductance is approximately independent of bias in our samples, as well.  At biases beyond 300 mV, our ST-FMR measurements suggest that the torkance might increase strongly, whereas the parallel conductance does not, but we suspect that this apparent torkance increase may be an artifact (perhaps related to hot-electron effects or heating [65]).  One reason for this suspicion is that for |V| > 300 mV we also observe increases in the resonant linewidth that are inconsistent with the simple macrospin spin-torque theory, which suggests that bias-dependent effects in addition to spin-torque become important in this regime.  The Tsukuba-ANELVA-Osaka group found an in-plane torkance that was approximately independent of bias for |V| < 100 mV, but which then increased strongly in magnitude at larger negative biases and decreased in magnitude and even changed sign at positive bias.  The differences between the data of the two collaborations at high bias lead us to suspect even more strongly that one should consider carefully how hot-electron effects and other potential artifacts might influence ST-FMR experiments when large *dc* biases are applied.

The proportionality between the in-plane torkance and the parallel conductance predicted by Equations (4) and (5) is expected to be accurate only when the cross-conductances $G_{+-}$ and $G_{-+}$ remain equal, so that they cancel in Eq. (4).  This cancellation seems reasonable for tunnel junctions with the symmetric structure that has been measured thus far (CoFeB electrodes on both top and bottom), but we would expect that $G_{+-}$ and  $G_{-+}$ might differ when $V{\neq}0$ for more asymmetric devices.  This could produce an in-plane torque with a much stronger bias dependence, more similar to the strong bias dependence seen in the antiparallel conductance.  Future studies of this sort of device might be very interesting, because the form of this bias dependence might provide a way to distinguish between the different mechanisms that have been proposed to affect the bias dependence of the conductance -- magnon excitation, impurity effects, density-of-states variations in the electrodes, and Coulomb correlation effects [57,70].

Heating in MTJs

One of the important practical differences between spin-transfer effects in MTJs and all-metal multilayers is heating.  In all-metal devices that are optimized to give low critical currents for spin-transfer excitations, the degree of heating is generally negligible.  For an applied current of 1.0 mA, about twice the critical current for spin-torque-driven precession in one 5-Ω all-metal device tested [71], the effective magnetic temperature in the device rose to about 20 K from 4.2 K, which (by Eq. (2) in ref. [71], $T_{eff} = \sqrt{T_0^2 + \beta I^2}$, with $\beta$ a sample-dependent constant) should correspond to a heating at room temperature of only 0.7 K.  Tunnel junctions experience temperature increases that are a great deal larger, because their greater resistances produce much more Ohmic heating, and because tunneling electrons may experience additional inelastic scattering processes.  In MTJs with resistances greater than 1 kΩ, a current bias of just 0.2 mA has been found to give a change in effective temperature for the magnetic free layer $\Delta T > 30$ K, starting at room temperature [69,72]. For a bias of 1 mA, this extrapolates to $\Delta T$ greater than 400 K at room temperature.

There are a number of interesting open questions associated with heating in MTJs, with implications both for basic science and applications. One important question is whether the effective temperature of the magnet will be determined by simple Ohmic heating alone, or whether the relatively high biases applied to tunnel junctions may excite short-wavelength spin-wave excitations directly by means of inelastic spin-flip scattering. This could conceivably produce an even higher effective magnetic temperature, distinct from any effective temperature of the conduction electrons in the device. Because of the far-from-equilibrium conditions present for an MTJ under bias, it is also an open question whether the magnetic excitations would be well-described by an effective thermal distribution, or whether some more sophisticated approach might be needed to understand their properties. For practical applications, the effect of heating is likely to be mostly negative -- for instance, the added randomness from increased thermal fluctuations may impact the reproducibility of switching in memory devices. However, there may be situations in which heating can be harnessed for positive effect. For example, if heating can be used to quickly decrease the total magnetic moment in a nanoscale layer, this might make a given spin-transfer torque more effective in reversing the moment, and provide a mechanism to reduce the critical current for switching.

Decreasing the Critical Current for Spin-Transfer-Driven Excitations in MTJs

Much of the applied research relating to spin-transfer torques in tunnel junctions in the near future will likely be oriented toward two goals: decreasing the critical current (or voltage) needed to produce reversal in memory devices, and increasing the reliability of the junctions by improving their resistance to damage under high bias. Both goals are necessary for making nonvolatile magnetic memory devices in which information is written using the spin-transfer effect. Currently, one needs biases close to the levels that produce dielectric breakdown in tunnel barriers in order to produce ns-scale switching in realistic memory bits. Because Katine and Fullerton have contributed a separate article about applications of the spin-transfer effect, we will comment only very briefly about these issues here. However, we will note that the results of the ST-FMR experiments discussed above provide some guidance as to what strategies for reducing switching currents are likely to be most fruitful.

We explained above that the ST-FMR measurements show that the in-plane component of the spin-transfer torque in CoFeB/MgO/CoFeB tunnel junctions has (within an experimental uncertainty of 15%) a magnitude in quantitative agreement with the prediction given by Eq. (6). The devices on which these measurements were made had a TMR of 154%, corresponding to a tunneling spin polarization $P_S$ = 0.66. According to the prediction in Eq. (6), if one were to increase the quality of the tunnel junction to give perfect polarization ($P_S$ = 1), the ratio of the in-plane torque to the parallel differential conductance would increase by only about 8%. In this sense, the existing CoFeB/MgO/CoFeB tunnel junctions are close to being ideal -- one should expect to gain very little extra torque strength by improving the polarization for a tunnel junction of a given parallel conductance. Research efforts for decreasing critical currents for magnetic switching should therefore be focused on other strategies. These can include using more complicated multilayer structures so as to apply spin-transfer torques to both sides of a switching layer [72,73], reducing the magnetic damping, decreasing the magnetization of the switching layer to reduce its total angular momentum [74], applying oscillatory signals to drive magnetic layers resonantly [75], and/or using materials with perpendicular rather than in-plane magnetic anisotropy [76]. It is also worth further effort to investigate whether the significant perpendicular component of the spin-torque in tunnel junctions can be harnessed to decrease switching currents or improve switching times. To date, with only a few exceptions [77], simulations of spin-transfer switching have not considered the consequences of having a torque component in this direction.

Other Topics That Interest Us

In closing this article, instead of looking back and summarizing what has been accomplished in the past, we will attempt to look forward at a few topics that may come to be increasingly interesting in the future.

*The angular dependence of spin torque*: The spin-transfer torque is predicted to depend on the angle $\theta$ between $\vec{M}_R$ and $\vec{M}_L$ differently for tunnel junctions and metal spin valves. For tunnel junctions, a simple $\sin(\theta)$ dependence is predicted [51], while for metal spin valves the torque strength is expected to be asymmetric about $\theta = \pi/2$, with the degree of asymmetry related to deviations in the resistance from a simple $(1-\cos(\theta))$ dependence [60]. Measuring this dependence would likely be a very sensitive test of the theories of spin torque. There are also puzzling questions associated with apparent asymmetries in spin-torque phenomena between initially parallel and initially antiparallel magnetic orientations -- switching currents and *dc*-driven precessional dynamics appear to be more asymmetric in some devices than in others, for reasons that are not clear. Being able to measure the angular dependence of the torque accurately for both configurations might help to explain why. Furthermore, measurements of spin torque as a function of angle might also allow a determination of the magnetic damping as a function of angle, and a means to distinguish between different mechanisms which can produce magnetic damping. Thus far we are aware of one measurement of spin torque *vs.* $\theta$ in a metal spin valve structure, based on the critical currents for *dc*-driven precessional excitations [78], and no detailed study yet in tunnel junctions. Additional research along these lines would certainly be interesting.

*Very thin magnetic layers*: We mentioned above that the derivation of the spin-torque in tunnel junctions assumes that a spin current flowing within a magnetic layer does not have a component perpendicular to the magnetic moment. This is an approximation. If one injects electrons into a magnetic layer with spins tilted at an angle with respect to the magnetic moment, they will precess in the exchange field of the magnet. The spins traveling along different trajectories will precess by different angles and will therefore become effectively randomized after a certain distance of penetration, and it is only after this point that the net transfer of transverse angular momentum from the current to the magnet can be considered complete, so that the spin current will not have an average component perpendicular to the magnet's moment. The working assumption for tunnel junctions has been that the angular momentum transfer occurs in a very thin interface region, of the order of an atomic layer or two, adjacent to the tunnel barrier. This assumption, however, is just a simple extension from the model for a metallic multilayer. We do not yet know quantitatively, either experimentally or theoretically, the thickness dependence of the spin torque exerted on the free layer in a tunnel junction. One way to explore this might be to examine spin transfer in very thin magnetic layers, to look for new phenomena in free layers that might be thinner than a minimum thickness needed to completely absorb the transverse spin current [79]. It will be interesting to see if this distance may differ for hot electrons at high bias *vs*. those close to the Fermi energy, or for tunnel junctions in which tunneling is highly focused in the direction perpendicular to the barrier *vs*. junctions with tunneling over a broader spread of angles.

*A proper micromagnetic understanding*: An increasingly important issue, for spin-transfer devices made from either metal multilayers or tunnel junctions, is to understand the full dynamics of the magnetic layers with correct micromagnetic simulations. Spin-transfer experiments are often analyzed with simple macrospin models that neglect spatial variations in the magnetic dynamics within layers. This has been done to achieve a reasonable zeroth-order understanding of the dynamics, and in fact in many cases the macrospin models have provided descriptions and predictions that are surprisingly accurate. However, in reality, deviations from spatially-uniform dynamics are probably important, to varying degrees, in all spin-transfer devices, and a rigorous micromagnetic approach will be necessary for a full understanding. This is particularly clear in ST-FMR studies, because they observe signals from spatially non-uniform spin-wave modes, in addition to larger signals from a mode that appears to be approximately spatially uniform [64,65]. It will be interesting to see if extensions of the ST-FMR technique may be able to examine energy flow from the more-uniform fundamental mode to the spatially non-uniform modes due to their non-linear coupling, as this is likely one of the fundamental mechanisms giving magnetic damping. Micromagnetic simulations will also be unquestionably vital for tunnel devices designed so that current flow (and therefore the spin transfer torque) are not uniform through the cross section of the device. Initial measurements have suggested that a non-uniform current flow may be an effective strategy for reducing the critical current required for magnetic switching [80].

*Spin-transfer excitations of ferrimagnets and antiferromagnets*: Spin-torque effects have been studied thus far primarily with simple transition-metal magnets, with a few experiments on ferromagnetic semiconductor devices [81]. It will be interesting for several reasons to examine ferrimagnets and antiferromagnetic materials, as well. One possible strategy for decreasing the critical current for switching in memory devices is to decrease the total magnetic moment of the free layer, so that less angular momentum transfer may be required to get it to switch. If ferrimagnets provide a way to decrease the total magnetic moment while maintaining thermal stability and without increasing the magnetic damping, they could be very important for technology. The IBM group has begun to explore the ferrimagnet CoGd with some of these ideas in mind [82]. In regard to antiferromagnets, MacDonald and collaborators have predicted that it may be possible to manipulate the order parameter in antiferromagnets in ways that are analogous to the spin-transfer excitations that have been demonstrated in ferromagnets, with possibly even lower switching currents [83,84]. (See also the article by Haney, Duine, Núñez, and MacDonald in this volume.) There are initial experimental indications that a spin-polarized current can be used to adjust the strength of the exchange bias between antiferromagnetic and ferromagnetic layers [85,86]. If it becomes possible to excite antiferromagnetic spin resonances with spin-transfer torques, similar to past results with ferromagnetic normal modes, this might enable the production of nanoscale oscillators and resonators with frequencies reaching to 100's of GHz.

*New phenomena in nanoscale magnetic tunnel junctions?*: Spin-torque experiments to date have been performed with devices having cross sections on the order of 100 nm or wider, for which the transfer of angular momentum and the resulting magnetic dynamics can be understood within an essentially classical-physics picture. This scenario should change in nanoparticles [87] or single magnetic molecules [88,89] at low temperature, where Coulomb-blockade effects and the quantum mechanics of tunneling via discrete electronic levels must be taken into account. The new phenomena that might be observed in this regime have begun to be explored theoretically [90,91,92,93,94], but not yet experimentally. One way in which nanoparticles and magnetic molecules might be qualitatively different from larger devices is that the time scale for relaxation of magnetic excitations could be much longer, because the discrete spectrum may inhibit spin relaxation. Whereas 100-nm-scale magnets have characteristic relaxation times on the order of nanoseconds, two initial experiments on magnetic nanoparticles [87,95] have set lower bounds on the relaxation time for spin excitations of 1 microsecond and 150 ns. If these limits are correct, the critical scale of current density required for spin-transfer switching or dynamical excitations might be reduced greatly compared to measurements of larger devices, in inverse proportion to the relaxation time.

Acknowledgements: We thank John Slonczewski and Bob Buhrman for discussions and comments and Sufei Shi for help with figures. We also acknowledge fruitful discussions with Stuart Parkin, and thank Y. Nagamine, D. D. Djayaprawira, N. Watanabe, and K. Tsunekawa of Canon ANELVA Corp. for providing the junction thin film for the data shown in Fig.1. The work of DCR is supported by the Office of Naval Research, the National Science Foundation (DMR-0605742 and EEC-0646547), and DARPA.

around the direction of the magnetization vector as the electron moves through the magnet, due to the exchange field. In the transition-metal magnets Fe, Ni, Co and their alloys, the exchange interaction is sufficiently strong that in one period of precession an electron can move only a short distance -- a few atomic lengths. Therefore, if two electrons start with the same initial spin direction but travel along even slightly different paths within a magnet, then the transverse components of their spin orientations (transverse to the magnet's magnetization vector) will quickly get out of phase. As a consequence, when a current of spin-polarized electrons enters into a ferromagnet, any initial spin component transverse to the magnetization should quickly become dephased and average to zero, so that well inside a ferromagnet the current cannot have a spin component in any direction other than parallel or antiparallel to the magnetization.